\documentclass[twocolumn,showpacs,preprintnumbers,draftclsnofoot]{revtex4}
\usepackage{amsmath}
\usepackage{amssymb}
\usepackage{graphicx}
\usepackage{dcolumn}
\usepackage{bm}
\usepackage{amssymb}
\usepackage{graphicx}
\usepackage{dcolumn}
\usepackage{bm}

\begin{document}

\title{ Topological zero-line mode of bilayer graphene with Rashba spin-orbital coupling and staggered sublattice potentials  }
\author{Ma Luo\footnote{Corresponding author:luom28@mail.sysu.edu.cn} and Zhibing Li}
\affiliation{The State Key Laboratory of Optoelectronic Materials and Technologies \\
School of Physics\\
Sun Yat-Sen University, Guangzhou, 510275, P.R. China}

\begin{abstract}

Domain wall in bilayer graphene with Rashba spin-orbital coupling and staggered sublattice potentials, at the interface between two domains with different gated voltages, is studied. Varying type of zero-line modes are identified, including zero-line mode with pure spin filtering effect. The Y-shape current partition at the junction among three different domains are proposed.

\end{abstract}

 \maketitle

\section{Introduction}

Bilayer graphene(BLG) has tunable band gap that can be controlled by the gate voltage\cite{Eduardo07}. When gated voltages of opposite signs are applied on  the left and right domains of the BLG, the zero-line modes(ZLMs) that are localized near to the interface and exponentially decaying away from the interface are presented\cite{Ivar08,Matthew11,Zarenia11,Klinovaja12,Abolhassan13,FanZhang13,XintaoBi15,ChangheeLee16}. The dispersions are chiral, i.e. the group velocities at K and K' valley are opposite to each other. Current partition among four ZLMs at a junction has been discussed \cite{ZhenhuaQiao11,ZhenhuaQiao14,Anglin17,KeWang17,YafeiRen17}.

With the present of sufficiently large Rashba spin-orbital coupling(SOC)\cite{HongkiMin06,Fufang07,Zarea09,Rashba09,Zhenhua10,Klinovaja12,Santos16}, the gated BLG become topological insulator\cite{qiao11,qiao13}. The present of staggered sublattice potentials at top and(or) bottom graphene layer changes the phase diagram  significantly\cite{Xuechao16,MaLuo18}. There are four phases, which are  the quantum spin Hall(QSH) phase with topological invariant $Z_{2}=1$ and valley Chern number  $C_{V}=2$, the quantum valley Hall(QVH) phase with $Z_{2}=0$ and $C_{V}=4$, band insulator(BI) phase with $Z_{2}=0$ and $C_{V}=0$, and edge conductive metal(EM) phase with $Z_{2}=0$ and $C_{V}=0$. As comparison, the suspended BLGs have only QSH and QVH phases.

In this paper, ZLMs of the gated BLG with Rashba SOC and staggered sublattice potentials along the zigzag edge are studied. The gated voltages of  two adjoining domains are not necessarily opposite to each other. Rich phases of the topological ZLMs are identified. A new current partition scheme is proposed. In section II, the ZLMs with the absence of the Rashba SOC is reviewed and extended to the case with staggered sublattice potentials. In section III, the ZLMs with the present of the Rashba SOC are studied. The ZLMs in two BLGs with and without staggered sublattice potentials are shown as examples. In section IV, we proposed a scheme of Y-shape current partition. In section V, the conclusion is given.

\section{BLGs with staggered sublattice potential}

The BLG is modeled by continue Dirac Fermion model, which gives the Hamiltonian as
\begin{equation}
H_{A}=\left[\begin{array}{cccc}
V(x)-\Delta_{1} & \hat{p} & 0 & 0 \\
\hat{p}^{+} & V(x)+\Delta_{1} & t_{\perp} & 0 \\
0 & t_{\perp} & -V(x)-\Delta_{2} & \hat{p} \\
0 & 0 & \hat{p}^{+} & -V(x)+\Delta_{2} \\
\end{array}\right]\label{Hamiltonian}
\end{equation}
where $\hat{p}=-i\hbar v_{F}(\tau\partial_{x}-i\partial_{y})$ and $\hat{p}^{+}=-i\hbar v_{F}(\tau\partial_{x}+i\partial_{y})$ with $v_{F}=c_0/330$ being the Fermi velocity of graphene, $c_0$ being the speed of light; $t_{\perp}=0.39eV$ is the inter-layer hopping term between two overlapping carbons, one from the   top and the other from the  bottom layers; $\Delta_{1}$ and $\Delta_{2}$ are the staggered sublattice potentials for bottom and top graphene layers respectively; $V(x)$ is the potential difference induced by the gated voltage; $\tau=\pm1$ for K and K' valleys respectively. The zigzag edge is along the y axis with $x=0$. A kink potential at $x=0$ with $V(x\gtrless0)$ being constant are used throughout this paper. The smooth step-like potential induces additional edge modes near to the bulk band edge, but the ZLMs remain robust\cite{Ivar08,XintaoBi15}.

In each domain, the plane wave solution of the four component spinor is $|\psi\rangle=\chi e^{ik_{x}x+ik_{y}}$, with $\chi=[\chi_{1A},\chi_{1B},\chi_{2A},\chi_{2B}]^{T}$. Four components stand for the amplitudes at the A and B sub-lattices of the 1st and 2nd layers respectively. The system has translational symmetry along y axis, so that $k_{y}$ is a good quantum number. Only the ZLMs with energy level $\varepsilon$ being within the bulk gap are studied, whose $k_{x}$ always has non-zero imaginary part. For the spatially localized ZLMs, the imaginary part of $k_{x}$ is positive(negative) in the domain with $x>0$($x<0$). Inserting the plane wave solution into the eigen equation $H_{A}|\psi\rangle=\varepsilon|\psi\rangle$ and analytically solving the equation, one finds the relations between $k_{x}$ and $\varepsilon$ to be $(\hbar v_{F}k_{x})^2=-\Delta^2+\varepsilon^2+V^2-(\hbar v_{F}k_{y})^2\pm\sqrt{(\varepsilon^2-\Delta^{2})t_{\perp}^{2}+(4\varepsilon^2-t_{\perp}^{2})V^{2}-2\Delta Vt_{\perp}^{2}}$ for $\Delta_{1}=\Delta_2=\Delta$, and $(\hbar v_{F}k_{x})^2=-\Delta^2+\varepsilon^2+V^2-(\hbar v_{F}k_{y})^2\pm\sqrt{(\varepsilon^2+\Delta^{2})t_{\perp}^{2}+(4\varepsilon^2-t_{\perp}^{2})V^{2}-2\Delta \varepsilon t_{\perp}^{2}}$ for $\Delta_{1}=-\Delta_2=\Delta$. Note that although the dispersion relations only depend on $V^2$, $V(x<0)$ and $V(x>0)$ have different magnitude in our study, therefore $k_{x}$ for each domain needs to be  calculated  separately. With a given pair of $\varepsilon$ and $k_{y}$, and the corresponding $k_{x}$, the normalized spinor $\chi$ is obtained, either analytically or numerically, by finding the null space of the matrix $H_{A}-\varepsilon I$ with $I$ being four by four unit matrix. Among the four solutions of $k_{x}$, only two of them with proper sign of imaginary part is chosen for each domain. Thus, the wave functions for $x>0$ and $x<0$ domain are
\begin{equation}
|\Psi\rangle^{\gtrless}=(u_{1}^{\gtrless}\chi_{1}^{\gtrless}e^{ik_{x}^{1,\gtrless}x}+u_{2}^{\gtrless}\chi_{2}^{\gtrless}e^{ik_{x}^{2,\gtrless}x})e^{ik_{y}y} \label{wavefunction}
\end{equation}
At $x=0$, all four components of the wave functions are continuous, giving four matching equations for the four unknown constants $u_{1,2}^{\gtrless}$. The ZLMs are found by searching the pairs of $\varepsilon$ and $k_{y}$ that the determinant of the coefficient  matrix of the matching equations being zero. With $\Delta_{1}=\Delta_2=0$, the ZLMs have been given in the previous publication \cite{Ivar08}.

\begin{figure}[tbp]
\scalebox{0.36}{\includegraphics{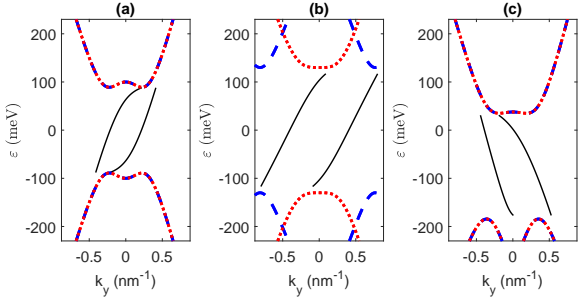}}
\caption{ Band structure of the ZLMs of BLG with (a) $\Delta_1=\Delta_2=0$, $V(x<0)=-V(x>0)=100meV$, (b) $\Delta_1=\Delta_2=130meV$, $V(x<0)=0$, $V(x>0)=468meV$, and (c) $\Delta_1=-\Delta_2=130meV$, $V(x<0)=-V(x>0)=168meV$. The bulk band edge at $x<0$ and $x>0$ domains are plotted as blue(dash) and red(dotted) lines, respectively. }
\label{fig_band1}
\end{figure}

Numerical results of three types of ZLMs are presented in Fig. \ref{fig_band1}. The ZLMs in suspended BLGs in figure (a) is the same as known. We consider the stagger lattice potential induced by $SiC$ substrate\cite{SYZhou07}. For the BLGs with $\Delta_1=\Delta_2=\Delta=130meV$, the bulk gap is close at $V=\Delta$, and open with $2\Delta$ at $V=0$. With the gated voltage being $V=468meV$, the bulk gap is also $2\Delta$. We studied the ZLMs at the domain wall with $V=0$($V=468meV$) at the left(right) domain. Figure (b) shows that the bulk bands of two domains are different, but with the same gap. The dispersion of the ZLMs is nearly linear for a wide range of energy. For the BLGs with $\Delta_1=-\Delta_2=130meV$ in figure (c), the particle-hole symmetry is broken, so that the band structure is not symmetric about $\varepsilon=0$. All presented ZLMs are in the K valley. The band structure of the corresponding ZLMs in the K' valley can be obtained by mirror reflection $k_y\rightarrow-k_y$ of that in the K valley.

\section{BLGs with Rashba SOC}

With the Rashba SOC, the Hamiltonian becomes\cite{Julien14,Inglot14,Inglot15,luoma17}
\begin{equation}
H=\left[\begin{array}{cc}
H_{A} & H_{R} \\
H_{R}^{+} & H_{A} \\
\end{array}\right]\label{HamiltonianR}
\end{equation}
where
\begin{equation}
H_{R}=i\hbar\Omega_{R}\left[\begin{array}{cccc}
0 & -\tau+1 & 0 & 0 \\
-\tau-1 & 0 & 0 & 0 \\
0 & 0 & 0 & -\tau+1 \\
0 & 0 & -\tau-1 & 0 \\
\end{array}\right]
\end{equation}
and $\hbar\Omega_{R}$ is the strength of the Rashba SOC. The plane wave solution is spinor having  eight components, $|\psi\rangle_{R}=\chi_{R} e^{ik_{x}x+ik_{y}}$, with $\chi_{R}=[\chi_{1A\uparrow},\chi_{1B\uparrow},\chi_{2A\uparrow},\chi_{2B\uparrow},\chi_{1A\downarrow},\chi_{1B\downarrow},\chi_{2A\downarrow},\chi_{2B\downarrow}]^{T}$ where the arrows stand for spin up and down.
The relation between $k_{x}^{2}$ and $\varepsilon$ is a quartic equation, whose analytical solution is too lengthy to be written in the paper. According to the same principle described in the last section, four solutions of $k_{x}$ out of eight with proper sign of the imaginary part are chosen for each domain. The wave function  for each domain is a supposition of four plane waves, which is similar to Eq. (\ref{wavefunction}). The matching condition at $x=0$ gives  a system of eight linear equations. The  ZLMs are found by searching for the pairs of $\varepsilon$ and $k_y$ that corresponding to the zero point of the determinant of the coefficient  matrix.

\begin{figure}[tbp]
\scalebox{0.58}{\includegraphics{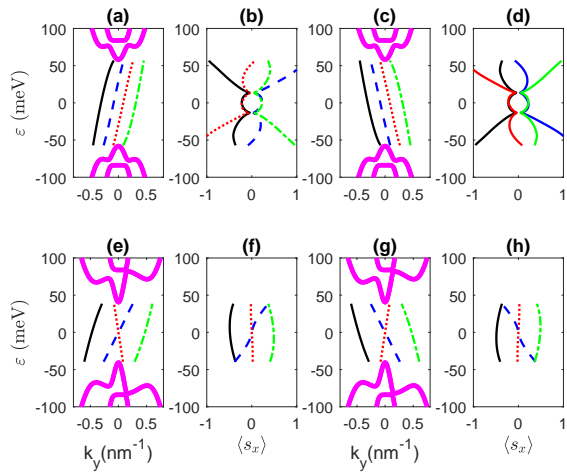}}
\caption{ The first and third columns are band structures of the ZLMs in K and K' valleys, respectively, and the corresponding expectation  of spin x versus energy level are in the second and fourth column. The systems are the suspended BLGs with $V(x<0)=-V(x>0)=83.8meV$, and with $\hbar\Omega_{R}=84meV$ in (a-d), $\hbar\Omega_{R}=336meV$ in (e-h).  The bulk band edges are plotted as thick lines. The bands of ZLMs and the expectations  of spin x of ZLMs are plotted as thin lines. Different ZLM is represented by different line styles. }
\label{fig_band2}
\end{figure}

We firstly investigate the ZLMs in the suspended BLG. The results with $V(x\gtrless0)=\pm V$ are shown in Fig. \ref{fig_band2}. When the strength of Rashba SOC is small, the system is in the trivial insulator phase. The SOC  splits each ZLM into two bands as shown in Fig. \ref{fig_band2}(a-d). The dispersions of ZLMs are chiral, meaning that the group velocities in the K and K' valley are opposite to each other, and the spin expectation in K and K' valley have the same sign. It can be seen by comparing each thin line of the first column with the thin line of the same style of the third column in Fig. \ref{fig_band2}. The results of sufficient large  strength of the Rashba SOC ($\hbar\Omega_R=336meV$) is shown in Fig. \ref{fig_band2}(e-h), where  the BLG is driven  into the topological insulator phase. For the two ZLMs being plotted as black(solid) and green(dash-dot) lines, the dispersions are chiral with the spin expectation in K and K' valley having the same sign, which is similar to the ZLMs in trivial insulator. For the other two ZLMs being plotted as red(dotted) and blue(dashed) lines, the spin expectation in K and K' valley are opposite to each other. The later two ZLMs are originated from the band inversion of the bulk band of the topological insulator.

\begin{figure}[tbp]
\scalebox{0.5}{\includegraphics{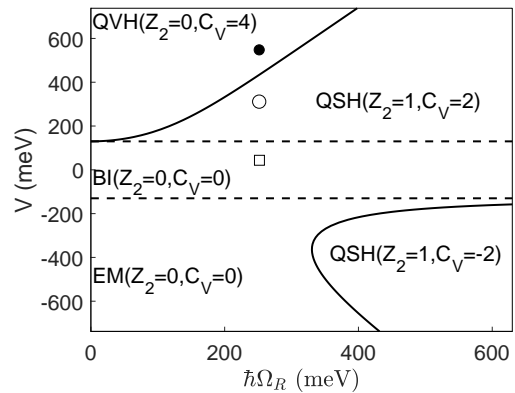}}
\caption{ Phase diagram of the BLGs with $\Delta_{1}=\Delta_{2}=130meV$.  }
\label{fig_phase}
\end{figure}

\begin{figure}[tbp]
\scalebox{0.58}{\includegraphics{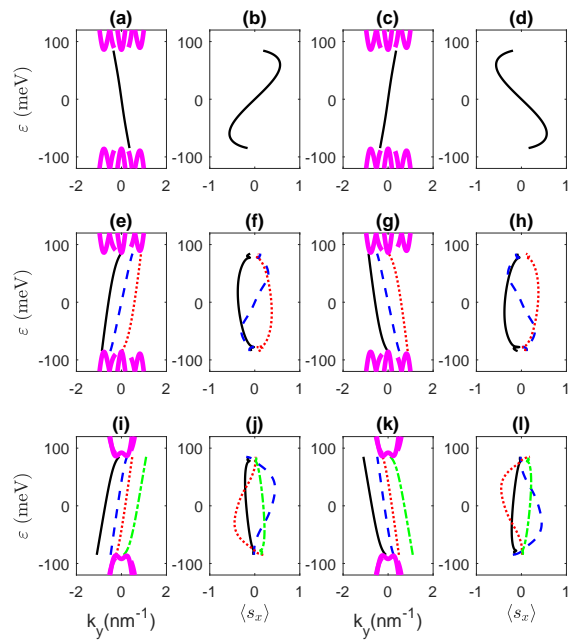}}
\caption{ The same type of plotting as Fig. \ref{fig_band2} for the BLGs with $\Delta_{1}=\Delta_{2}=130meV$, $\hbar\Omega_{R}=252meV$. (a-d) for ZLM-I with $V(x<0)=310meV$ and $V(x>0)=540meV$; (e-h) for the ZLM-II with $V(x<0)=310meV$ and $V(x>0)=43meV$; (i-l) for the ZLM-III with $V(x<0)=540meV$ and $V(x>0)=43meV$. }
\label{fig_band3}
\end{figure}

Secondly, we study the BLGs with substrates being $\Delta_{1}=\Delta_{2}=130meV$. The phase diagram is plotted in Fig. \ref{fig_phase} in the space of $\hbar\Omega_{R}$ and $V$. We are particularly interested in the BLGs with $\hbar\Omega_{R}=252meV$ and $V$ being one of $546meV$, $310meV$ and $43meV$. The three systems, which are marked respectively by dot, circle and square in Fig. \ref{fig_phase}, are in topological phases QVH, QSH and BI, respectively, while all have the same band gap. The domain wall between any two of them supports a ZLM. We denote the ZLM at the domain wall between QSH and QVH as ZLM-I, that between QSH and BI as ZLM-II, and that between QVH and BI as ZLM-III. The band structure and spin expectation of the ZLM-I are plotted in Fig. \ref{fig_band3}(a-d). There is only one band with nearly perfectly linear dispersion. The dispersions are chiral. At the same energy level, the spin expectation  in K and K' valley are opposite to each other. Thus, the elastic inter-valley back scattering of the ZLM is completely forbidden. The ZLM-I supports pure spin filtering effect. The band structure and spin expectation  of the ZLM-II are plotted in Fig. \ref{fig_band3}(e-h). For the two bands plotted by black(solid) and red(dotted) lines, the spin expectation at the same energy level in K and K' valley have the same sign, so that they are trivial chiral bands. The band plotted as blue(dashed) line has nearly linear dispersion, and opposite spin expectation in K and K' valley, so that this band has spin filtering effect. ZLM-I and ZLM-II are at the domain wall between the topological trivial and non-trivial insulators. ZLM-III is at the domain wall between two topological trivial insulators with different valley Chern numbers. The band structure and spin expectation of the ZLM-III are plotted in Fig. \ref{fig_band3}(i-l). All of these four bands are trivial chiral bands.

\section{Y-shape current partition}

\begin{figure}[tbp]
\scalebox{0.5}{\includegraphics{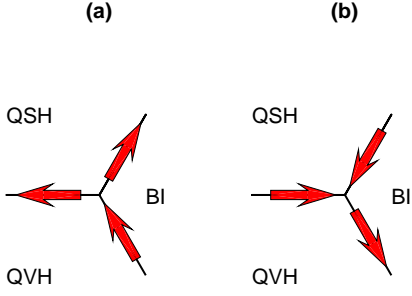}}
\caption{ The Y-shape current partition at the junction of three ZLMs. The directions of group velocities along the domain walls  at K and K' valleys are  shown in (a) and (b) respectively as  indicated by the arrows.    }
\label{fig_partition}
\end{figure}

The current partition scheme at a suspended BLG with $V$ switching sign requires even number of ZLMs at the junction\cite{ZhenhuaQiao11,ZhenhuaQiao14,Anglin17,KeWang17,YafeiRen17}. We show in the following that a Y-shape current partition is possible when staggered potentials and Rashba SOC are involved. For the BLG with $\Delta_{1}=\Delta_{2}=130meV$ and $\hbar\Omega_{R}=252meV$, which is investigated in the last section, the gated voltage can drive the BLG into three different topological phases  as shown in Fig. \ref{fig_phase}. Thus, the junction of three phases allows current partition among three ZLMs, as shown in Fig. \ref{fig_partition}. The group velocity of one ZLM in K and K' valley are plotted in Fig. \ref{fig_partition}(a) and (b), respectively. Assuming the absence of inter-valley scattering, the incident current from the ZLM-III is partitioned into ZLM-I and ZLM-II, as shown in Fig. \ref{fig_partition}(a). On the other hand, the incident current from the ZLM-I or ZLM-II is redirected to ZLM-III, without current partition as shown in Fig. \ref{fig_partition}(b). The Y-shape junction induces small inter-valley scattering, producing small back scattering into the incident ZLM. In addition, due to the difference of number of channels, as well as the difference of the wave number and transverse wave functions of the three ZLMs, the transmission coefficients from the import ZLM to the outport ZLMs are not integer and the current partition is not evenly distributed.

\begin{figure}[tbp]
\scalebox{0.21}{\includegraphics{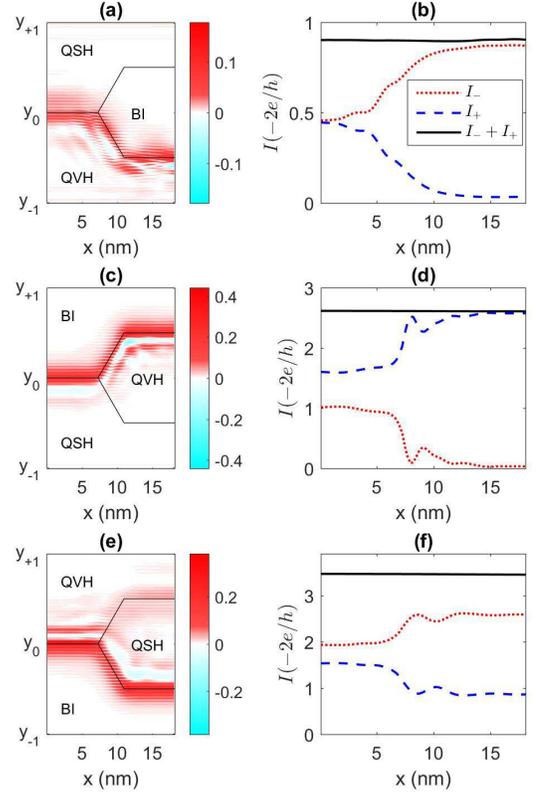}}
\caption{ Current distribution of Y-shape junction with import from ZLM-I, ZLM-II and ZLM-III in figure (a,b), (c,d) and (e,f), respectively. The 2D spatial distributions of $j_{x}$ are plotted in the left column, with the black(solid) lines being the domain walls. $I_{-}$, $I_{+}$ and $I_{-}+I_{+}$ are plotted as red(dotted), blue(dash) and black(solid) lines in the right column. }
\label{fig_current}
\end{figure}

The proposed scheme for current partition is confirmed by calculations of transportation based on the Landauer-Buettiker formalism and recursively constructed Green's functions of the tight binding model\cite{LopezSancho84,Nardelli98,FufangXu07,Diniz12,Lewenkopf13}. The conductance between two leads is numerically calculated from $G_{pq}=(e^{2}/h)Tr[\Gamma_{p}G^{r}\Gamma_{q}G^{a}]$, with $(e^{2}/h)$ being elementary conductance, $G^{r}$ and $G^{a}$ being retarded and advance Green's functions of the scattering region, $\Gamma_{p}$ and $\Gamma_{q}$ being the line width matrixes of the lead p and q, respectively. For the non-partition current, conductance from ZLM-I to ZLM-III is $0.26(e^{2}/h)$, conductance from ZLM-II to ZLM-III is $1.74(e^{2}/h)$. For the partition current, conductance from ZLM-III to ZLM-I is $0.24(e^{2}/h)$, conductance from ZLM-III to ZLM-II is $1.69(e^{2}/h)$. The local current between two lattice sites is calculated as
\begin{equation}
\mathbf{j}(\mathbf{r}_{i}\rightarrow\mathbf{r}_{j})=-\frac{2e\hat{\mathbf{d}}_{ij}}{h}\int dE[t_{ij}G^{<}(\mathbf{r}_{j},\mathbf{r}_{i})-t_{ji}G^{<}(\mathbf{r}_{i},\mathbf{r}_{j})]
\end{equation}
where $\mathbf{r}_{i}$ is the position of the i-th lattice site, $\hat{\mathbf{d}}_{ij}$ is the unit vector from the i-th to j-th lattice site, $e$ is the electron charge, $h$ is the Planck constant, $t_{ij}$ is the hopping parameter between the i-th and j-th lattice site, and $G^{<}(\mathbf{r}_{i},\mathbf{r}_{j})$ is the lesser Green's function. The currents through the cross section $(y_{-1},y_{0})$ and $(y_{0},y_{+1})$ as a function of x coordinate are defined as $I_{-}=\int_{y_{-1}}^{y_{0}}j_{x}dy$ and $I_{+}=\int_{y_{0}}^{y_{+1}}j_{x}dy$. The 2D spatial distributions of $j_{x}$ and the cross section currents are plotted in Fig. \ref{fig_current}(a,b), (c,d) and (e,f) with imports from the ZLM-I, ZLM-II and ZLM-III, respectively. At the import side, the incident current are evenly distribute into $I_{-}$ and $I_{+}$. Due to the spatially asymmetric reflection at the Y-shape junction, $I_{-}$ and $I_{+}$ at the import side are different from each other. With import from the ZLM-I, the reflection is suppressed as shown in Fig. \ref{fig_current}(a) and (b), because of the pure spin filtering effect of the ZLM-I. At the output side, the current is redirect into $I_{-}$ or $I_{+}$ in Fig. \ref{fig_current}(b) or (d), respectively; the current is partitioned into $I_{-}$ and $I_{+}$ unevenly in Fig. \ref{fig_current}(f). These phenomenons agree with the partition rule of the Y-shape junction. The total cross section current $I_{-}+I_{+}$ remains constant due to the charge conservation.

\section{conclusion}

In summary, the ZLMs in the BLGs with Rashba SOC and staggered sublattice potentials are investigated. The domain wall between two regions with different gated voltages supports localized chiral edge modes with varying numbers of channels. Particular ZLM with only one channel exhibiting nearly perfectly linear dispersion and pure spin filtering effect is identified. Diversified ZLMs with different dispersions and spin expectation can be obtained by engineering the domain wall between different topological phases. As an example, for the BLGs with $\Delta_{1}=\Delta_{2}$ and fixed $\hbar\Omega_{R}$, tuning the gated voltage into three values drives the BLGs into three different topological phases with the same band gap. It is shown that the  Y-shape junction of the three different topological phases with odd number of domain walls is allowed. The current partition rule at the Y-shape junction is investigated. The diversified ZLMs and Y-shape current partition could boost the development of integrated opto-spintronic and valleytronic devices.

\begin{acknowledgments}
The project is supported by the National Natural Science Foundation of China (Grant:
11704419), the National Basic Research
Program of China (Grant: 2013CB933601), and the National Key Research and Development Project of China
(Grant: 2016YFA0202001).
\end{acknowledgments}

\clearpage

\end{document}